\documentclass{article}

\usepackage[preprint]{neurips_2023}

\usepackage[utf8]{inputenc} 
\usepackage[T1]{fontenc}    
\usepackage{hyperref}       
\usepackage{url}            
\usepackage{booktabs}       
\usepackage{amsfonts}       
\usepackage{nicefrac}       
\usepackage{microtype}      
\usepackage{xcolor}         
\usepackage{longtable}
\usepackage{setspace}
\usepackage{etoolbox}
\AtBeginEnvironment{quote}{\singlespace\vspace{-\topsep}}
\AtEndEnvironment{quote}{\vspace{-\topsep}\endsinglespace}

\title{A Legal Risk Taxonomy for\\Generative Artificial Intelligence}

\author{
    David Atkinson\hspace{1.0em}
    Jacob Morrison
    \AND \textnormal{Allen Institute for Artificial Intelligence}
\\
 \small{
   \textbf{Correspondence:} {\{davida, jacobm\}@allenai.org}
 }
}

\begin{document}

\maketitle

\begin{abstract}

\renewcommand*{\thefootnote}{\fnsymbol{footnote}}
\footnotetext{The views and opinions expressed in this paper are those of the authors and do not necessarily reflect the official policy or position of their employers.}
\renewcommand*{\thefootnote}{\arabic{footnote}}

For the first time, this paper presents a taxonomy of legal risks associated with generative AI (GenAI) by breaking down complex legal concepts to provide a common understanding of potential legal challenges for developing and deploying GenAI models. The methodology is based on (1) examining the legal claims that have been filed in existing lawsuits and (2) evaluating the reasonably foreseeable legal claims that may be filed in future lawsuits. First, we identified 29 lawsuits against prominent GenAI entities and tallied the claims of each lawsuit. From there, we identified seven claims that are cited at least four times across these lawsuits as the most likely claims for future GenAI lawsuits. For each of these seven claims, we describe the elements of the claim (what the plaintiff must prove to prevail) and provide an example of how it may apply to GenAI. Next, we identified 30 other potential claims that we consider to be more speculative, because they have been included in fewer than four lawsuits or have yet to be filed. We further separated those 30 claims into 19 that are most likely to be made in relation to pre-deployment of GenAI models and 11 that are more likely to be made in connection with post-deployment of GenAI models since the legal risks will vary between entities that create versus deploy them. For each of these claims, we describe the elements of the claim and the potential remedies that plaintiffs may seek to help entities determine their legal risks in developing or deploying GenAI. Lastly, we close the paper by noting the novelty of GenAI technology and propose some applications for the paper’s taxonomy in driving further research.

\end{abstract}

\section{Introduction}

Generative artificial intelligence (GenAI)\footnote{GenAI is not to be confused with artificial general intelligence (AGI), which is not-yet-existent broad human-level intelligence.} is a form of artificial intelligence (AI) that can create new outputs, such as new text, images, or audio. GenAI models have expanded from single-modality (e.g., text or image-based)\footnote{e.g., ChatGPT, Stable Diffusion} to multi-modal (e.g., generating both images and text, and possibly audio, code, and video)\footnote{e.g., Gemini}. More importantly, GenAI models are being used in many different contexts spanning medical application \citep{singhal2023expertlevel} to companionship\footnote{e.g., Replika, PI}, legal review\footnote{e.g., Harvey}, and virtually every form of creative work\footnote{e.g., Firefly, MusicLM}.

The benefits of GenAI are not without their risks, however. Researchers have already identified a slew of potential risks and categorized them into groups and hierarchies called taxonomies, such as ethical \citep{taxonomyfromdeepmind}, sociotechnical \citep{weidinger2023sociotechnical}, code generating \citep{khlaaf2022hazard}, privacy \footnote{Shaping the future: A dynamic taxonomy for AI privacy risks: \href{https://iapp.org/news/a/shaping-the-future-a-dynamic-taxonomy-for-ai-privacy-risks/}{https://iapp.org/news/a/shaping-the-future-a-dynamic-taxonomy-for-ai-privacy-risks/}}, and security \citep{gupta2023chatgpt, Derner2023ASR} risks. There are also specific taxonomies of risks for text-to-image \citep{bird2023typology}, audio generation \citep{Barnett_2023}, and other modality-specific use cases.

Most, if not all, of the taxonomies cite actual harms, as opposed to merely theorizing about potential harms. There are also databases that keep a running catalog of real-world harms caused by GenAI\footnote{\href{https://incidentdatabase.ai/}{AI Incident Database}} \footnote {\href{https://oecd.ai/en/incidents?search\_terms=\%5B\%5D\&and\_condition=false\&from\_date=2014-01-01\&to\_date=2024-02-29\&properties\_config=\%7B\%22principles\%22:\%5B\%5D,\%22industries\%22:\%5B\%5D,\%22harm\_types\%22:\%5B\%5D,\%22harm\_levels\%22:\%5B\%5D,\%22harmed\_entities\%22:\%5B\%5D\%7D\&only\_threats=false\&order\_by=date\&num\_results=20}{AIM: The OECD AI Incidents Monitor, an evidence base for trustworthy AI}}.

Notably missing, however, is a taxonomy focused on the legal risks associated with developing and deploying GenAI despite nearly two dozen lawsuits against the likes of Meta, Anthropic, OpenAI, and Alphabet with claims that vary widely in legal theory, such as copyright, torts, contracts, unfair competition, trademark dilution, and more. 

A legal taxonomy is important to distinguish from others given the particular risks and harms that apply in the legal context. For example, GenAI that hypersexualizes women\citep{Wolfe_2023}, hallucinates\citep{xu2024hallucination}, or spouts harmful disinformation\citep{chen2023combating} may have tremendous ethical implications that may cause significant reputational damage to the developers and deployers of those systems. However, it’s unlikely such ethical issues will cause the type of existential harm that losing a class action lawsuit might. Indeed, many prominent GenAI models have already produced most ethical harms known to be practically possible (not merely theoretical) and yet the models and their creators still exist and, in many cases, still generate more revenue or achieve a higher market cap each month than most businesses.

This paper provides the first comprehensive overview of GenAI legal risks\footnote{To be clear, this paper is not legal advice. It is an overview from a legal perspective. If you need legal advice, you should hire a lawyer.}. Definitions for legal terminology used throughout this paper are included in Table~\ref{tab:definitions}.

\begin{table}[]
\begin{tabular}{| p{0.23\linewidth} | p{0.73\linewidth} |}
\hline
\multicolumn{1}{|c|}{\textbf{Term}} & \multicolumn{1}{c|}{\textbf{Definition}} \\ \hline
\multicolumn{1}{|l|}{Case law} & Case law, also referred to as common law, is a law that is based on precedents set by courts, rather than law as written in constitutions, statutes, or regulations. \\ \hline
\multicolumn{1}{|l|}{Civil damages} & Civil damages (or damages) are monetary awards that the plaintiff can seek from the defendant in a civil lawsuit. The amount and type of damages depend on the nature and extent of the harm, and the applicable law. \\ \hline
\multicolumn{1}{|l|}{Civil law} & Civil law includes actions in which the parties bringing the suits (the plaintiffs) are seeking to enforce private obligations or duties against the other parties (the defendants). \\ \hline
\multicolumn{1}{|l|}{Criminal law} & Criminal law is based on written statutes by which a state or the federal government prohibits specific kinds of conduct. The statutes allow the government to impose fines or imprisonment on persons convicted of violating them. Criminal cases are always brought by the government whose law has allegedly been violated. \\ \hline
\multicolumn{1}{|l|}{Criminal penalties} & These are fines or imprisonment that the government can impose on the defendant for violating a criminal statute. The penalties under these laws can vary depending on the intent, harm, and prior convictions of the defendant. \\ \hline
\multicolumn{1}{|l|}{Common law} & Unlike criminal law, which is based on statutes, common law is the body of law created by judges in written opinions (the rulings of the court). Common law courts look to the past decisions of courts to synthesize the legal principles of past cases. \\ \hline
\multicolumn{1}{|l|}{Compensatory damages} & These damages aim to restore the plaintiff to the same situation he or she was in before the injury occurred. Compensatory damages can cover both tangible and intangible losses, such as medical expenses, lost wages, property damage, pain and suffering, emotional distress, and loss of enjoyment of life. \\ \hline
\multicolumn{1}{|l|}{Defendant} & An individual, company, or institution sued or accused in a court of law. \\ \hline
\multicolumn{1}{|l|}{\textit{de novo}} & \textit{De novo} is a Latin term that means "anew," "from the beginning," or "afresh." When a court hears a case “\textit{de novo},” it is deciding the issues without reference to any legal conclusion or assumption made by a previous court. \\ \hline
\multicolumn{1}{|l|}{Element/Factor} & An element (sometimes called a factor) is an essential requirement to make a claim or defense in court. \\ \hline
\multicolumn{1}{|l|}{Injunction} & Injunctions require the person to stop doing something or to perform a specific action to remedy the harm. Injunctions may be temporary or permanent, depending on the circumstances of the case. \\ \hline
\multicolumn{1}{|l|}{Liability} & The state of being responsible for something \\ \hline
\multicolumn{1}{|l|}{Nominal damages} & These damages recognize that the plaintiff was legally wronged, but did not suffer any actual harm. \\ \hline
\multicolumn{1}{|l|}{Remedy} & A form of court enforcement of a legal right resulting from a successful civil lawsuit that includes awarding damages or ordering an injunction. \\ \hline
\multicolumn{1}{|l|}{Statute} & A written law passed by a legislative body. Often just called a law. \\ \hline
\multicolumn{1}{|l|}{Plaintiff} & A person who brings a case against another in a court of law. \\ \hline
\multicolumn{1}{|l|}{Private right of action} & Allows an individual or organization (rather than just the government) to bring a lawsuit in court based on an alleged violation of a law and to seek a remedy. \\ \hline
\multicolumn{1}{|l|}{Prosecution / Prosecutor} & The legal party responsible for presenting the case in a criminal trial against the defendant, an individual accused of breaking the law. \\ \hline
\multicolumn{1}{|l|}{Punitive damages} & Aims to punish the defendant for his or her malicious, reckless, or fraudulent conduct, and to deter others from engaging in similar behavior. \\ \hline
\end{tabular}
\caption{Glossary of terms and definitions related to GenAI legal risks.}
\label{tab:definitions}
\end{table}

\section{Scope}

There are (unfortunately) dozens of potential legal claims that could conceivably be brought against GenAI developers and deployers depending on the particular facts of the situation, and how adventurous and creative the plaintiffs would like to be in bringing lawsuits. Indeed, because GenAI is so new, the body of case law applied to it is vanishingly small. For the most part, it remains unsettled as to how existing statutes and common law will apply to what may be a sufficiently unique form of technology to merit review de novo. Some of the claims below may ultimately be dismissed in most instances, but we include them in this paper for completeness. In short, the actual legal risk for certain common claims is unknown.

Though the nature of novel legal theories is no doubt riveting, this paper only aspires to discuss at some length those claims that have already been brought in federal court against prominent GenAI entities, and are therefore probably the most likely claims to be brought in future litigation against other GenAI entities over at least the next few years. We provide a more cursory overview of less likely or less common claims.

\section{The Most Litigated Claims}

This section will discuss the claims that are most often cited in existing GenAI cases based on a simple methodology: we tallied the frequency of each claim that appears in 29 lawsuits against prominent GenAI entities\footnote{These are the only lawsuits based on GenAI claims the authors are aware of. There may be more.}. Note that frequency does not necessarily equate to strength. Just because a certain legal claim occurs more frequently than another does not mean the claim is more likely to succeed. However, frequency is a good proxy for the types of arguments that plaintiffs believe have at least a reasonable chance of success. We identified seven claims that are cited most frequently across these lawsuits as further discussed below: direct copyright infringement, vicarious copyright infringement, contributory copyright infringement, DMCA 1202(b), unjust enrichment, negligence, and unfair competition. 

\subsection{Copyright Infringement}

\subsubsection{Direct Copyright Infringement}

Direct copyright infringement is when a person, without authorization (e.g., in the absence of a license or other agreement between the parties), reproduces, distributes, displays, or performs a copyrighted work, or prepares a derivative work based on a copyrighted work as these are exclusive rights of the copyright owner\footnote{17 U.S. Code § 106}. Notably, copyright infringement applies strict liability, meaning it does not require knowledge of infringement or intent to infringe by the defendant. In other words, the plaintiff does not need to prove that the defendant knowingly or intentionally made unauthorized use of the plaintiff’s copyrighted work in order to prevail. 

How does one know if there was infringement? Courts use a standard called “substantial similarity,” which is a way of comparing the alleged infringing material to the original to decide whether they are too similar\footnote{\href{https://plus.lexis.com/api/permalink/7ace1710-76cc-4b8a-b375-62681d1d78c9/?context=1530671}{https://plus.lexis.com/api/permalink/7ace1710-76cc-4b8a-b375-62681d1d78c9/?context=1530671}}. The mechanics of how courts conduct the assessment varies by jurisdiction, but the underlying principle of comparing one version to another remains unchanged. One might wonder why such a test is required, and the reason is because if only exact copies of the original counted as infringement, people would need to only make the slightest alteration to escape a copyright claim. 

Infringement is not the end of the story, however. In some circumstances, the unauthorized use of copyrighted works is considered fair use. Fair use is a legal doctrine that permits the use of copyrighted works without authorization from the copyright owner and exempts one from liability for copyright infringement using a four-part test  to determine whether something qualifies as fair use, as summarized below.

\begin{quote}
Section 107 of the Copyright Act\footnote{\href{https://www.law.cornell.edu/uscode/text/17/107}{17 U.S. Code § 107 - Limitations on exclusive rights: Fair use}} identifies four factors to consider when assessing whether the use of copyrighted material is infringement or fair use:
\begin{enumerate}
    \item The purpose and character of the use, including whether such use is of a commercial nature or is for nonprofit educational purposes;
    \item The nature of the copyrighted work;
    \item The amount and substantiality of the portion used in relation to the copyrighted work as a whole; and
    \item The effect of the use upon the potential market for or value of the copyrighted work.
\end{enumerate}
\end{quote}

The fair use test requires an assessment of all the factors together. The courts have repeatedly emphasized that there are no bright line rules, and that each case must be decided on its own facts\footnote{As of the publication of this paper there is no case law on the application of fair use to GenAI specifically.}.

Direct copyright infringement could apply to GenAI at several steps of the GenAI development and deployment pipeline, including during data collection, model training, when the model is released, and at the point when outputs are generated based on a prompt. Most legal scholars seem to believe that inputs to the model (i.e., the training materials and training process) are probably fair use, while outputs tend to be less defensible\footnote{See, e.g., \href{https://texaslawreview.org/fair-learning/}{https://texaslawreview.org/fair-learning/}}.

The penalties for direct infringement can include statutory damages ranging from \$750 to \$30,000 per work, or up to \$150,000 per work for willful infringement, as well as actual damages and profits, injunctions, and attorney’s fees\footnote{\href{https://www.copyright.gov/title17/chapter5.pdf}{https://www.copyright.gov/title17/chapter5.pdf}}.

\subsubsection{Vicarious Copyright Infringement}

Vicarious infringement is a form of secondary liability for copyright infringement created by common law. It allows for Party A to be found liable for the infringing acts of Party B if (i) Party A had the the right and ability to control the infringing activity and (ii) Party A had a direct financial interest in the infringement. Importantly there must have been direct infringement by Party B for Party A to possibly face secondary infringement.

This form of liability may apply to GenAI entities if they, for example, host a model that is producing infringing content and they make people pay to use the model. By hosting the model, the GenAI entity probably has the ability to apply guardrails that would prevent infringing outputs, possibly satisfying the first prong, while the payment they receive may satisfy the second prong.

The penalties for vicarious infringement can include statutory damages ranging from \$750 to \$30,000 per work, or up to \$150,000 per work for willful infringement, as well as actual damages and profits, injunctions, and attorney’s fees\footnote{\href{https://www.copyright.gov/title17/chapter5.pdf}{https://www.copyright.gov/title17/chapter5.pdf}}.

\subsubsection{Contributory Copyright Infringement}

Contributory copyright infringement is the second type of secondary liability commonly claimed in ongoing litigation. A successful claim requires that (i) Party A makes a material contribution to the infringing activity, while (ii) having knowledge or a reason to know of the direct infringement by Party B. Making a material contribution could include knowingly inducing Party B to directly infringe, for example\footnote{\href{https://plus.lexis.com/api/permalink/80c8265b-c026-4196-8dbb-9091b8646790/?context=1530671}{https://plus.lexis.com/api/permalink/80c8265b-c026-4196-8dbb-9091b8646790/?context=1530671}}.

Contributory infringement may apply to GenAI if someone hosts a model that they know can  produce infringing content (e.g., images of Nintendo characters) and someone tells the model host that such infringing content is being produced by their model, but the host takes no action to mitigate the infringement. 

The penalties for contributory infringement can include statutory damages ranging from \$750 to \$30,000 per work, or up to \$150,000 per work for willful infringement, as well as actual damages and profits, injunctions, and attorney’s fees\footnote{\href{https://www.copyright.gov/title17/chapter5.pdf}{https://www.copyright.gov/title17/chapter5.pdf}}.

\subsubsection{Digital Millennium Copyright Act 1202(b)}

Section 1202(b) of the Digital Millennium Copyright Act (DMCA) was meant to prohibit people from removing or altering copyright management information (sometimes called CMI, which includes information like the creator’s name and title of the work). It requires the intent to remove the CMI, distributing CMI knowing that the CMI has been removed or altered without authorization,  or distributing copies of works knowing the CMI has been removed or altered, provided that the person knew or had reason to know, that such removal or alteration of CMI would “induce, enable, facilitate, or conceal an infringement…”\footnote{\href{https://www.law.cornell.edu/uscode/text/17/1202}{17 U.S. Code § 1202 - Integrity of copyright management information}} Because this law requires knowledge or intent of both the removal or alteration of CMI and that such removal or alteration would lead to infringement, it, like most laws that require knowledge or intent, is difficult to prove. 

Section 1202(b) could apply where an entity collects copyrighted works, then deletes the CMI prior to training a model, and then generates infringing outputs that don’t contain the CMI. However, the case law on this matter seems to require that the output be an exact copy of the original work, which is another high bar to meet\footnote{\href{https://storage.courtlistener.com/recap/gov.uscourts.cand.403220/gov.uscourts.cand.403220.195.0_1.pdf}{https://storage.courtlistener.com/recap/gov.uscourts.cand.403220/gov.uscourts.cand.403220.195.0\_1.pdf}}.

The penalties for violating the section 1202 of the DMCA can include the actual damages and any additional profits of the violator, or at any time before final judgment is entered, a complaining party may elect to recover an award of statutory damages for each violation of section 1202 in the sum of not less than \$2,500 or more than \$25,000\footnote{\href{https://www.law.cornell.edu/uscode/text/17/1203}{17 U.S. Code § 1203 - Civil remedies}}.

\subsection{Unjust Enrichment}

Unjust enrichment is common law and as such is subject to different requirements in different states. Generally, it occurs when Party A is in some agreement with Party B. Party A confers a benefit on Party B, but Party B does not comply with the terms of the agreement yet still retains the benefit conferred by Party A\footnote{\href{https://www.law.cornell.edu/wex/unjust\_enrichment}{https://www.law.cornell.edu/wex/unjust\_enrichment}}. Some plaintiffs try to broaden the scope of this claim, though, to situations where there is no agreement, such that Party B receives a benefit from Party A without Party A’s assent or consent. 

The second scenario above is the one that will most likely apply to GenAI. For example, if a GenAI developer uses Party A’s books as training material to create the GenAI without Party A’s consent, and then the GenAI developer makes money from hosting the model, Party A may claim that revenue was the ill-gotten unjust enrichment.

There are two types of remedies for unjust enrichment: personal remedies and proprietary remedies.
\begin{enumerate}
    \item A \textit{personal} remedy is when the court orders the unjustly enriched party to pay the monetary value of the benefit they received to the other party. This is also called restitution, and it is the most common remedy for unjust enrichment\footnote{\href{https://content.next.westlaw.com/practical-law/document/Ib9aa1b161c9a11e38578f7ccc38dcbee/Restitution?viewType=FullText\&transitionType=Default\&contextData=(sc.Default)}{https://content.next.westlaw.com/practical-law/document/Ib9aa1b161c9a11e38578f7ccc38dcbee/Restitution}}.
    \item A \textit{proprietary} remedy is when the court orders the unjustly enriched party to return the specific property or asset that they gained to the other party. This is also called disgorgement, and it is usually granted when the property or asset is unique or irreplaceable\footnote{\href{https://www.law.cornell.edu/wex/disgorgement\#:\~:text=Disgorgement\%20is\%20a\%20remedy\%20requiring,As\%20seen\%20in\%20SEC\%20v}{https://www.law.cornell.edu/wex/disgorgement}}.
\end{enumerate}

\subsection{Negligence}

Negligence is a tort, which is a body of law that fills in the void left by contract law and criminal law where someone is harmed and someone else must be liable even if the action that caused the harm wasn’t in breach of a contract or in violation of a criminal law. Negligence claims must satisfy four elements\footnote{\href{https://www.law.cornell.edu/wex/damages}{https://www.law.cornell.edu/wex/damages}}: 

\begin{enumerate}
    \item the liable party must have owed a duty to act as a reasonably prudent person would under the circumstances, 
    \item the liable party breached that duty by acting unreasonably, 
    \item the breach of duty was the cause of the harm to the injured party, and 
    \item the injured party was in fact injured\footnote{\href{https://www.law.cornell.edu/wex/negligence}{https://www.law.cornell.edu/wex/negligence}}.
\end{enumerate}

Application of negligence to GenAI can vary widely. Almost anytime someone feels injured, they can work their way back to blaming someone for the harm. One negligence claim, for example, could hinge on whether the GenAI developers acted reasonably when curating its training dataset to not collect and use third-party trade secrets. Another claim could center on whether reasonable precautions and guardrails were put in place to prevent or mitigate the model from providing harmful information to someone who then uses that information to cause harm. 

The prevailing party can receive civil damages. Those damages can fall into one of the following:
\begin{enumerate}
    \item \textit{Nominal damages}: These are symbolic awards that recognize that the plaintiff was legally wronged, but did not suffer any actual harm. Nominal damages are rare in negligence cases, as negligence usually requires proof of injury.
    \item \textit{Compensatory damages}: These are awards that aim to restore the plaintiff to the same situation he or she was in before the negligence occurred. Compensatory damages can cover both tangible and intangible losses, such as medical expenses, lost wages, property damage, pain and suffering, emotional distress, and loss of enjoyment of life. 
    \item \textit{Punitive damages}: These are awards that aim to punish the defendant for his or her malicious, reckless, or fraudulent conduct, and to deter others from engaging in similar behavior. Punitive damages are not available in all jurisdictions, and are subject to certain limits and standards.
\end{enumerate}

\subsection{Unfair Competition}

Most claims of unfair competition are based on California law, which defines it as “any unlawful, unfair or fraudulent business act or practice and unfair, deceptive, untrue or misleading advertising…”\footnote{Cal. Bus. \& Prof. Code §§ 17200, et seq. (the “UCL”)} As one might expect, the statute is not precise enough on its own to make definitive conclusions about claims, so litigants must rely on case law to either bolster or distinguish their claim(s). 

This claim may apply to GenAI if, for instance, the GenAI developer promised a content creator that any data scraped from the content creator’s site would only be used for internal research purposes, but then subsequently used the material to offer a revenue-generating product to the public.

Generally, the remedies for unfair competition can be classified into two categories: monetary relief and injunctive relief\footnote{\href{https://www.findlaw.com/smallbusiness/business-laws-and-regulations/unfair-competition-.html\#:\~:text=Businesses\%20harmed\%20by\%20unfair\%20competition,in\%20state\%20and\%20federal\%20court}{https://www.findlaw.com/smallbusiness/business-laws-and-regulations/unfair-competition-.html}}. 
\begin{enumerate}
    \item \textit{Monetary} relief is when the court orders the violator to pay compensation to the injured party or to the government. This may include actual damages, lost profits, statutory damages, punitive damages, and attorney’s fees and costs. 
    \item \textit{Injunctive} relief is when the court orders the violator to stop or refrain from the unlawful conduct, or to take certain actions to correct or prevent the harm. This may include cease and desist orders, corrective advertising, product recall, disgorgement of profits, and compliance monitoring.
\end{enumerate}

\subsection{Mitigation Strategies}

The following non-comprehensive table identifies several ways GenAI developers and deployers can mitigate their legal risks pertaining to most of the possible claims discussed above. We separately explore potential mitigation strategies pre- and post-deployment/release of GenAI systems.

Pre-Deployment/Release:
\begin{enumerate}
    \item Use training material that:
        \begin{enumerate}
            \item Was collected by others 
            \item Was collected with consent
            \item Is not from sites you should know harbor illegally-obtained datasets 
            \item Doesn’t include “unpublished” works\footnote{\href{https://www.oyez.org/cases/1984/83-1632}{https://www.oyez.org/cases/1984/83-1632}}
            \item Is public domain 
            \item Is properly-licensed\footnote{Beware: \href{https://spectrum.ieee.org/data-ai}{https://spectrum.ieee.org/data-ai}}
            \item Was gathered by respecting the instructions of robots.txt, and/or 
            \item Doesn’t use the deep web or dark web for data
        \end{enumerate}
    \item Filter the training data to remove material that might encode harmful behaviors.
    \item Remove duplicates from the training set to reduce the likelihood of the model memorizing any given portion of training material
    \item When releasing datasets, release in non-human-readable formats (preferably in formats that are irreversible to human-readable without great effort and technical know-how).
    \item Remove PII/personal data.
    \item Train the model(s) with strong levels of differential privacy, which not only protects PII, but also makes it so virtually nobody can distinguish, with high probability, between a model trained with a particular training example and one trained without it \citep{neel2024privacy}.
    \item Conduct extensive red-teaming to try to coax potential illegal outputs from the model prior to publicly releasing it, then develop ways to minimize such occurrences in the future. 
    \item Use reinforcement learning from human feedback (RLHF) to reduce the probability of copyright infringement, right of publicity violations, torts, and privacy-violating outputs.
    \item Invest in new mechanisms that make it difficult to fine-tune models for harmful uses, (e.g., “self destructing models” \citep{henderson2023selfdestructing})
    \item Don’t release or deploy models that pose a significant risk of violating the law.
    \item For text-to-image, remove non-public domain artist names associated with works so model users can’t generate outputs “in the style of” those artists.
    \item For text-to-image, replace copyrighted character names with more generic descriptors.
    \item Alternatively, change user prompts via prompt transformation (also called shadow prompting; covertly modifying the prompt on the back end) from specific artist or character names to more generic ones.
\end{enumerate}

Post-Deployment/Release:
\begin{enumerate}
    \item Release model(s) with responsible AI license (e.g., ImpACT or RAIL)
    \item Provide instance attribution if feasible.
    \item Compare model output to training set data to filter and block generations that implicate copyright or privacy.
    \item For live, deployed systems, create an easy way for users to report potentially harmful prompts or generations, and then continually update the prompt- and output-filtering tools
    \item Detect and filter out harmful and sensitive data that users provide.
    \item Implement watermarking or metadata tags on outputs.
    \item Limit access to research purposes if applicable.
    \item Allow a third-party audit and act on the recommendations (or record why action was not taken).
\end{enumerate}

\subsection{Open Questions}

For the sake of brevity, this paper will not discuss open questions at length. However, we do intend to host a webpage where we will explore unsettled legal questions in greater detail. What follows is merely a sampling of the types of legal questions around GenAI that we believe deserve greater attention.

\begin{enumerate}
    \item Does the provenance of data matter (e.g., should data intended to be publicly available be treated the same as data that is publicly available but wasn’t intended to be by the data creator?)
    \item How important is following robots.txt?
    \item How important, if at all, is allowing people to opt-out of datasets? Should this be balanced against the cost or technical burden of complying with an opt-out request?
    \item Must opt-out requests always be respected? If so, is it retroactive, or only for current datasets and future models?
    \item Do research organizations have to respect opt-out requests if doing so will make the research less reproducible? 
    \item How should licenses such as MIT or Apache, designed for code, apply to model weights?
    \item What is the threshold for sufficient information security to protect datasets and user inputs?
    \item Does the frequency of infringing outputs matter, or is a single output sufficient for infringement?
    \item Does it matter how many times a model must be prompted before it makes an infringing output?
    \item Should there be regulatory safe harbors or other legal incentives to promote transparency and safety?
    \item Is there a “reasonable effort” threshold/safe harbor for entities making good faith efforts to prevent offense, illicit, and infringing outputs?
\end{enumerate}

\section{Other Risks: Pre-deployment}

These claims either haven’t been made in GenAI litigation, so they’re more speculative, or they occur infrequently. Given their less prominent nature, this section will give them a more cursory overview. Here, we only provide a brief description of what the claim entails and the potential penalties if found liable or guilty so researchers can weigh the risks. 

\subsection{Copyright}

\subsubsection{DMCA 1201-1205}

Section 1202 of the Digital Millennium Copyright Act (DMCA) makes it unlawful to provide or distribute false copyright management information (CMI) with the intent to induce or conceal infringement. CMI is certain information, including the title, name of the author and copyright owner, and terms for use of the work, conveyed in connection with copies, phonorecords, performances, or displays of a work\footnote{\href{https://www.law.cornell.edu/uscode/text/17/chapter-12}{17 U.S. Code Chapter 12 - Copyright Protection and Management Systems}}.

The actual damages and any additional profits of the violator, or at any time before final judgment is entered, a complaining party may elect to recover an award of statutory damages for each violation of section 1202 in the sum of not less than \$2,500 or more than \$25,000\footnote{\href{https://www.law.cornell.edu/uscode/text/17/1203}{17 U.S. Code § 1203 - Civil Remedies}}.

\subsection{Contract}

\subsubsection{Breach of Licenses}

Most licenses, including permissive ones such as Apache 2.0, include requirements to: provide a copy of the license with any copies or derivatives; cause any modified files to carry prominent notices stating that the user changed the files; and the user must retain, in the source form of any derivative works that they distribute, all copyright, patent, trademark, and attribution notices from the source form of the work, excluding those notices that do not pertain to any part of the derivative works. If creators don't adhere to license terms yet use the code to train their model and provide code outputs, then they may be in breach of the license, which means both breach of contract and copyright infringement, to the extent copyright applies.

Potential penalties could include:
\begin{itemize}
    \item If an end user breaches a material provision of the End User License Agreement (EULA), the software publisher may have the right to terminate the agreement. 
    \item If the end user’s infringement causes financial harm to the software publisher or other parties, it could result in monetary penalties.
    \item In specific situations, EULA breaches may be regarded as criminal offenses and subject to criminal penalties such as fines or imprisonment.
    \item For willful infringement, statutory damages can be awarded up to \$150,000 per work infringed in copyright cases.
\end{itemize}

\subsubsection{Breach of Terms of Service}

Many websites have a Terms of Service or Terms of Use agreement and those terms may include prohibitions on how the site data can be accessed and/or used. There are several arguments for and against allowing liability for web scrapers based on only browsewrap and the law is unsettled when it comes to using scraped data to train LLMs.

The potential penalties are similar to breach of terms:
\begin{itemize}
    \item If you breach the ToS, the website operator may terminate your access to their services. 
    \item Violating the ToS might lead to civil lawsuits. For instance, if your actions cause financial harm to the website or other users, you could be held liable for damages.
    \item There have been cases where violating a website’s ToS was treated as a criminal offense. 
\end{itemize}

\subsection{Torts}

\subsubsection{Intrusion Upon Seclusion}

The laws regarding intrusion upon seclusion were established to protect any aspect of the plaintiff’s life that he can reasonably expect will not be intruded upon.  It’s one of the four privacy torts created under U.S. common law. 

The elements are generally: One who intentionally intrudes, physically or otherwise, upon the solitude or seclusion of another or his private affairs or concerns, is subject to liability to the other for invasion of his privacy, if the intrusion would be highly offensive to a reasonable person\footnote{\href{https://cyber.harvard.edu/privacy/Privacy\_R2d\_Torts\_Sections.htm}{https://cyber.harvard.edu/privacy/Privacy\_R2d\_Torts\_Sections.htm}}.

Civil penalties for intrusion upon seclusion can include compensatory and punitive damages. Nominal damages are rare in intrusion cases, as they usually require proof of injury.

In some jurisdictions intrusion upon seclusion can also be considered a criminal offense under the laws of trespassing, voyeurism, stalking, harassment, or cybercrime. 

\subsubsection{Tortious Interference with a Prospective Economic Advantage}

Tortious interference with a prospective economic relationship occurs when a party improperly prevents two others from entering a contract or otherwise doing business together. Interference with prospective advantage does not require proof of a legally binding contract. 

The elements are generally: 
\begin{enumerate}
    \item economic relationship between the plaintiff and some third party, with the probability of future economic benefit to the plaintiff, 
    \item the defendant’s knowledge of the relationship, 
    \item intentional acts on the part of the defendant designed to disrupt the relationship,
    \item actual disruption of the relationship, and
    \item economic harm to the plaintiff proximately caused by the acts of the defendant.
\end{enumerate}

Penalties can include civil damages. In some jurisdictions, such as Texas, tortious interference with a prospective economic relationship can additionally be considered a criminal offense under the laws of theft, fraud, or misappropriation.

\subsubsection{Tortious Interference with a Contract}

This is similar the the claim above, but the elements are slightly different, requiring: 
\begin{enumerate}
    \item the existence of a valid and enforceable contract between plaintiff and another, 
    \item defendant’s awareness of the contractual relationship, 
    \item defendant’s intentional and unjustified inducement of a breach of the contract, 
    \item a subsequent breach by the other caused by defendant’s wrongful conduct, and 
    \item damages\footnote{See, e.g., \href{https://www.law.cornell.edu/wex/intentional\_interference\_with\_contractual\_relations}{https://www.law.cornell.edu/wex/intentional\_interference\_with\_contractual\_relations}}.
\end{enumerate}

Penalties can include civil damages.

\subsubsection{Civil Conspiracy}

A civil conspiracy is a form of conspiracy involving an agreement between two or more parties to deprive a third party of legal rights or deceive a third party to obtain an illegal objective. A form of collusion, a conspiracy may also refer to a group of people who make an agreement to form a partnership in which each member becomes the agent or partner of every other member and engage in planning or agreeing to commit some act. It is not necessary that the conspirators be involved in all stages of planning or be aware of all details. Any voluntary agreement and some overt act by one conspirator in furtherance of the plan are the main elements necessary to prove a conspiracy. A conspiracy may exist whether legal means are used to accomplish illegal results, or illegal means used to accomplish something legal. 

The elements are:
\begin{enumerate}
    \item a combination of two or more persons, 
    \item the persons seek to accomplish an object or course of action, 
    \item the persons reach a meeting of the minds on the object or course of action, 
    \item one or more unlawful, overt acts are taken in pursuance of the object or course of action, and 
    \item damages occur as a proximate result\footnote{\href{https://www.findlaw.com/smallbusiness/business-laws-and-regulations/civil-conspiracy.html}{https://www.findlaw.com/smallbusiness/business-laws-and-regulations/civil-conspiracy.html}}.
\end{enumerate}

Penalties can include civil damages.

\subsubsection{Conversion}

If personal data is property (and some courts think it is), then by scraping it without permission the creator may be committing some form of conversion. For the purposes of conversion, “intent” merely means the objective to possess the property or exert property rights over it. A party is liable for conversion regardless of their knowledge of property’s ownership status.

The elements are:
\begin{enumerate}
    \item the plaintiff’s ownership or right to possession of the property, 
    \item the defendant’s conversion by wrongful act inconsistent with the property rights of the plaintiff, and 
    \item damages\footnote{See, e.g., \href{https://www.law.cornell.edu/wex/conversion}{https://www.law.cornell.edu/wex/conversion}}.
\end{enumerate}

Penalties can include civil damages.

\subsection{Property}

\subsubsection{Trespass to Chattels}

A plaintiff alleging trespass to chattels in the web scraping context generally must allege that a defendant accessed its computer system without authorization and caused damage. These claims often, but not always, turn on whether a plaintiff can allege that a scraper damaged its computer systems. Courts have differed on what constitutes such “damage.” In some cases, a plaintiff’s allegation of an increased burden on server capacity caused by a scraper’s activity may constitute damage for purposes of the claim.

The basic elements of a claim of trespass to chattels are: 
\begin{enumerate}
    \item the lack of the plaintiff's consent to the trespass,
    \item interference or intermeddling with possessory interest, and 
    \item the intentionality of the defendant's actions\footnote{See, e.g., \href{https://www.findlaw.com/injury/torts-and-personal-injuries/trespass-to-chattels.html}{https://www.findlaw.com/injury/torts-and-personal-injuries/trespass-to-chattels.html}}.
\end{enumerate}

In a trespass to chattel claim, you can typically recover actual damages (as opposed to nominal damages). Courts measure actual damages by assessing the diminished total value of the chattel resulting from the defendant’s actions. For instance, if someone damages your bicycle, the compensation would be based on the bike’s reduced value due to the damage.

\subsection{Criminal}

\subsubsection{Computer Fraud and Abuse Act (CFAA)}

The Computer Fraud and Abuse Act of 1986 (CFAA) is a United States cybersecurity bill. A claim under subsection 1030(a)(c) of the CFAA, for example, has three elements: 
\begin{enumerate}
    \item a defendant has accessed a protected computer; 
    \item has done so without authorization or by exceeding such authorization as was granted; and 
    \item he obtains information from the protected computer\footnote{\href{https://www.law.cornell.edu/uscode/text/18/1030}{18 U.S. Code § 1030 - Fraud and related activity in connection with computers}}.
\end{enumerate}

Penalties include fines per crime, imprisonment from 1 to 20 years, or a combination of both\footnote{\href{https://www.law.cornell.edu/uscode/text/18/1030}{18 U.S. Code § 1030 - Fraud and related activity in connection with computers}}.

\subsubsection{Aid and Abet}

Aiding and abetting is a legal doctrine related to the guilt of someone who aids or abets (encourages, incites) another person in the commission of a crime (or in another's suicide). It applies where it cannot be shown the party personally carried out the criminal offense, but where another person may have carried out the illegal act(s) as an agent of the charged, working together with or under the direction of the charged, who is an accessory to the crime. 

The elements are: 
\begin{enumerate}
    \item That the accused had specific intent to facilitate the commission of a crime by another; 
    \item the accused had the requisite intent of the underlying substantive offense; 
    \item the accused assisted or participated in the commission of the underlying substantive offense; and 
    \item someone committed the underlying offense\footnote{\href{https://www.justice.gov/archives/jm/criminal-resource-manual-2474-elements-aiding-and-abetting}{https://www.justice.gov/archives/jm/criminal-resource-manual-2474-elements-aiding-and-abetting}}.
\end{enumerate}

Penalties vary by state. For example, in Texas a first-degree felony, like theft of more than \$200,000 worth of property is punishable by five to 99 or life in prison, along with a \$10,000 fine\footnote{\href{https://statutes.capitol.texas.gov/Docs/PE/htm/PE.12.htm\#:\~:text=July\%2022\%2C\%202013.-,Sec.,or\%20less\%20than\%205\%20years}{https://statutes.capitol.texas.gov/Docs/PE/htm/PE.12.htm}}.

\subsubsection{Larceny/Theft}

If personal data is property (and some courts think it is) then by scraping it without permission the creator may be committing some form of property theft. 

The elements typically include:
\begin{enumerate}
    \item the unlawful taking, 
    \item of the property of another, 
    \item without their consent, and 
    \item with the specific intent to deprive the owner of that property. 
\end{enumerate}

As a more specific example, the California Penal Code Sec 496(a) applies to (i) every person who, (ii) buys or receives any property that has been stolen or that has been obtained in any manner constituting theft or extortion, (iii) knowing the property to be so stolen or obtained, or (iv) who conceals, sells, withholds, or aids in concealing, selling, or withholding any property from the owner, knowing the property to be so stolen or obtained.

Penalties will vary by state. For example, under California Penal Code Sec. 496 penalties could include: 
1) imprisonment, 
2) Any person who has been injured by a violation of subdivision (a) or (b) may bring an action for three times the amount of actual damages, if any, sustained by the plaintiff, costs of suit, and reasonable attorney's fees\footnote{\href{https://leginfo.legislature.ca.gov/faces/codes\_displaySection.xhtml?lawCode=PEN\&sectionNum=496}{https://leginfo.legislature.ca.gov/faces/codes\_displaySection.xhtml?lawCode=PEN\&sectionNum=496}}.

\subsection{Privacy}

\subsubsection{General Data Protection Regulation}

The European Union's General Data Protection Regulation (GDPR)\footnote{\href{https://gdpr-info.eu/}{https://gdpr-info.eu/}} applies to personal data related to EU residents, and covers data controllers, processors, and data subjects.

The GDPR (General Data Protection Regulation) is defined by its scope, rights of data subjects, and requirements imposed upon data processors:
\begin{itemize}
    \item Scope:
        \begin{itemize}
            \item Applies to personal data related to EU residents.
            \item Covers data controllers, processors, and data subjects.
        \end{itemize}
    \item Rights of Data Subjects:
        \begin{itemize}
            \item Right to Access: Individuals can request information about their data processing.
            \item Right to Rectification: Correct inaccuracies in personal data.
            \item Right to Erasure (Right to Be Forgotten): Request deletion of personal data.
            \item Right to Data Portability: Obtain and reuse personal data for other services.
            \item Right to Object: Object to processing based on legitimate interests.
        \end{itemize}
    \item Legal Basis for Processing:
        \begin{itemize}
            \item Requires a lawful basis (e.g., consent, contract, legal obligation).
        \end{itemize}
    \item Data Protection Officer (DPO):
        \begin{itemize}
            \item Appoint a DPO if processing is regular or systematic.
        \end{itemize}
    \item Breach Notification:
        \begin{itemize}
            \item Notify authorities within 72 hours of a data breach.
        \end{itemize}
    \item Privacy by Design and Default:
        \begin{itemize}
            \item Integrate privacy into systems and processes.
        \end{itemize}
    \item Automated Decision-Making:
        \begin{itemize}
            \item Transparency and the right to human intervention.
        \end{itemize}

\end{itemize}

Fines for Non-Compliance: Up to 4\% of global annual turnover or €20 million, whichever is higher.

\subsubsection{California Consumer Privacy Act}

The California Consumer Privacy Act (CCPA)\footnote{\href{https://oag.ca.gov/privacy/ccpa}{https://oag.ca.gov/privacy/ccpa}} is similarly defined by its scope, consumer rights, and requirements imposed:
\begin{itemize}
    \item Scope:
        \begin{itemize}
            \item Applies to personal information of California residents.
            \item Covers businesses that meet specific criteria.
        \end{itemize}
    \item Consumer Rights:
        \begin{itemize}
            \item Right to Know: Request information about data collection.
            \item Right to Delete: Request deletion of personal information.
            \item Right to Opt-Out: Opt out of the sale of personal information.
            \item Right to Non-Discrimination: Businesses cannot discriminate based on exercising rights.
        \end{itemize}
    \item Transparency and Disclosure:
        \begin{itemize}
            \item Inform consumers about data collection practices.
        \end{itemize}
\end{itemize}

The California Attorney General enforces most of the CCPA.
Penalties could include fines for violations, but private right of action is limited to data breaches.
The private right of action only covers a data breach involving certain types of personal information.

\subsection{Miscellaneous}

\subsubsection{Data Broker Registration}

Two states currently have data broker registration laws.

California's data broker law requiring registration\footnote{\href{https://leginfo.legislature.ca.gov/faces/billNavClient.xhtml?bill\_id=202320240SB362}{https://leginfo.legislature.ca.gov/faces/billNavClient.xhtml?bill\_id=202320240SB362}}.
\begin{itemize}
    \item “Data broker” means a business that knowingly collects and sells to third parties the personal information of a consumer with whom the business does not have a direct relationship.
    \item California's law likely doesn't apply to nonprofits because they're not a "business.” 
\end{itemize}

Vermont's data broker law\footnote{\href{https://sos.vermont.gov/corporations/other-services/data-brokers/}{https://sos.vermont.gov/corporations/other-services/data-brokers/}}:
\begin{itemize}
\item  A Data Broker is a business, or unit or units of a business, separately or together, that knowingly collects and sells or licenses to third parties the brokered personal information of a consumer with whom the business does not have a direct relationship.
\item Vermont's law specifically says it doesn't matter if the entity was organized to operate at a profit.
\end{itemize}

Penalties could include: 
\begin{itemize}
    \item California: \$100 per day
    \item Vermont: \$50 per day, not to exceed \$10,000 per year for each year the entity fails to register
\end{itemize}

\subsubsection{Unfair or Deceptive Trade Practices}

According to the Federal Trade Commission (FTC)\footnote{\href{https://www.ftc.gov/about-ftc/mission/enforcement-authority}{https://www.ftc.gov/about-ftc/mission/enforcement-authority}}:

An act or practice is “unfair” if it “causes or is likely to cause substantial injury to consumers which is not reasonably avoidable by consumers themselves and not outweighed by countervailing benefits to consumers or to competition.”

“Deceptive” practices are defined in the Commission’s Policy Statement on Deception as involving a material representation, omission or practice that is likely to mislead a consumer acting reasonably in the circumstances.  

Civil Penalties\footnote{\href{https://www.law.cornell.edu/uscode/text/15/45}{15 U.S. Code § 45 - Unfair methods of competition unlawful; prevention by Commission}}:
\begin{itemize}
    \item The FTC can seek civil penalties against companies that engage in unfair or deceptive conduct. These penalties can exceed the profits earned through misconduct.
    \item Companies that receive a “Notice of Penalty Offenses” from the FTC and still engage in prohibited practices can face civil penalties of up to \$50,120 per violation\footnote{\href{https://www.ftc.gov/enforcement/penalty-offenses}{https://www.ftc.gov/enforcement/penalty-offenses}}.
\end{itemize}

Private Enforcement:
\begin{itemize}
    \item In some states, statutes allow private enforcement. 
    \item Individuals may be entitled to recover punitive damages and/or statutory fines.
\end{itemize}

State-Specific Penalties:
\begin{itemize}
    \item Each state has its own laws regarding deceptive trade practices. Penalties vary based on state statutes.
\end{itemize}

Injunctions and Restraint Orders:
\begin{itemize}
    \item Courts may issue injunctions or restraining orders to prevent continued deceptive trade practices.
\end{itemize}

\subsubsection{State Computer Access Laws}

State laws generally cover two areas:

\begin{enumerate}
\item Unauthorized Access and Computer Trespass:
\begin{itemize}
    \item Unauthorized access involves approaching, communicating with, or altering computer resources without consent.
    \item Laws in all 50 states, Puerto Rico, and the Virgin Islands address unauthorized access or computer trespass.
\end{itemize}

\item Malware and Viruses:
\begin{itemize}
    \item Malware and viruses modify, damage, destroy, or transmit information within computer systems without permission.
    \item Laws address actions that disrupt normal operation of computer systems.
\end{itemize}
\end{enumerate}

Penalties vary by state but may include fines, imprisonment, or both. For example:
\begin{itemize}
    \item In California, unlawful access, alteration, or damage to computer systems can result in fines, imprisonment, or both\footnote{\href{https://www.calpers.ca.gov/docs/ca-penal-code-502.pdf}{https://www.calpers.ca.gov/docs/ca-penal-code-502.pdf}}.
    \item In New Mexico, causing damage exceeding \$250 can lead to felony charges\footnote{\href{https://law.justia.com/codes/new-mexico/2021/chapter-30/article-45/section-30-45-5/\#:\~:text=Section\%2030\%2D45\%2D5\%20\%2D\%20Unauthorized\%20computer\%20use.\&text=E.,History\%3A\%20Laws\%201989\%2C\%20ch}{https://law.justia.com/codes/new-mexico/2021/chapter-30/article-45/section-30-45-5/}}.
\end{itemize}

\subsubsection{False Designation of Origin}

False designation of origin occurs when the manufacturer or seller lies about the country of origin or maker of its products.

Elements\footnote{\href{https://www.law.cornell.edu/uscode/text/15/1125}{15 U.S. Code § 1125 - False designations of origin, false descriptions, and dilution forbidden}}:
Any person who, on or in connection with any goods or services, or any container for goods, uses in commerce any word, term, name, symbol, or device, or any combination thereof, or any false designation of origin, false or misleading description of fact, or false or misleading representation of fact, which:
\begin{itemize}
\item is likely to cause confusion, or to cause mistake, or to deceive as to the affiliation, connection, or association of such person with another person, or as to the origin, sponsorship, or approval of his or her goods, services, or commercial activities by another person, or
\item in commercial advertising or promotion, misrepresents the nature, characteristics, qualities, or geographic origin of his or her or another person’s goods, services, or commercial activities, 
\end{itemize}
shall be liable in a civil action by any person who believes that he or she is or is likely to be damaged by such act.

Penalties could include\footnote{15 U.S.C. §1116(a)-§1118}:

\begin{itemize}
    \item Injunctions
    \item Any damages sustained by the plaintiff, defendant's profits, and the costs of the action.
    \item In exceptional cases, reasonable attorney fees
    \item The court may order that any infringing articles bearing the word, term, name, symbol, or device be destroyed
\end{itemize}

\subsubsection{Trademark Dilution}

Use in commerce of a mark or trade name if such use causes dilution by blurringe or tarnishment of the distinctive quality of a famous trademark\footnote{\href{https://www.everycrsreport.com/reports/RL34109.html\#\_Toc452631097}{https://www.everycrsreport.com/reports/RL34109.html\#\_Toc452631097}}.

Notes:
"Dilution" is statutorily defined in 15 U.S.C. §1127 to mean "the lessening of the capacity of a famous mark to identify and distinguish goods or services, regardless of the presence or absence of ... (1) competition between the owner of the famous mark and other parties, or (2) likelihood of confusion, mistake, or deception."

"Blurring" occurs when the famous mark's ability to identify its product has been impaired due to an association in the minds of consumers arising from similarity between another mark and the famous mark.

Tarnishment occurs when the reputation of a famous mark has been harmed by negative associations arising from the similarity between another mark and the famous mark. Situations in which tarnishment could result are when a famous trademark is "linked to products of shoddy quality, or is portrayed in an unwholesome or unsavory context, with the result that the public will associate the lack of quality or lack of prestige in the defendant's goods with the plaintiff's unrelated goods."

Penalties could include\footnote{Under 15 U.S.C. §1125(c)(5)}:
\begin{itemize}
    \item Injunctions.
    \item Owners of famous marks may also be entitled to the following additional remedies listed below if the mark or trade name that is likely to cause dilution by blurring or dilution by tarnishment was first used in commerce by the alleged infringer after Oct. 6, 2006; and (A) in a dilution by blurring action, the person willfully intended to trade on the recognition of the famous mark; or (B) in a dilution by tarnishment action, the person willfully intended to harm the reputation of the famous mark.
    \item For a willful violation, any damages sustained by the plaintiff, defendant's profits, and the costs of the action.
    \item In exceptional cases, reasonable attorney fees.
    \item For a willful violation, the court may order that any infringing articles bearing the word, term, name, symbol, or device be destroyed.
\end{itemize}

\subsubsection{Trade Secrets (State and Federal)}

The Uniform Trade Secrets Act (UTSA; which is broadly adopted by the states) defines a "trade secret" as:

information, including a formula, pattern, compilation, program, device, method, technique, or process that:
\begin{itemize}
    \item Derives independent economic value, actual or potential, from not being generally known to, and not being readily ascertainable by proper means by, other persons who can obtain economic value from its disclosure or use; and
    \item Is the subject of efforts that are reasonable under the circumstances to maintain its secrecy\footnote{\href{https://www.law.cornell.edu/wex/trade\_secret}{https://www.law.cornell.edu/wex/trade\_secret}}.
\end{itemize}

Remedies include both injunctive relief and monetary damages\footnote{\href{https://www.reuters.com/legal/legalindustry/remedies-trade-secret-misappropriation-2023-05-08/}{https://www.reuters.com/legal/legalindustry/remedies-trade-secret-misappropriation-2023-05-08/}}. In some instances, penalties could include imprisonment\footnote{\href{https://www.law.cornell.edu/uscode/text/18/1832}{18 U.S. Code § 1832 - Theft of trade secrets}}.

\section{Other Risks: Post-Deployment}

These claims either haven’t been made in GenAI litigation, so they’re more speculative, or they occur infrequently. Given their less prominent nature, this section, like the previous section, will give them a more cursory overview.

\subsection{Contract}

\subsubsection{Clickwrap and Enforceability of Artifact Terms}

Many websites have a Terms of Service or Terms of Use agreement and those terms may include prohibitions on how the site data can be accessed and/or used. There are several arguments for and against allowing liability for web scrapers based on only browsewrap and the law is unsettled when it comes to using scraped data to train LLMs. However, the consensus appears to be that only clickwrap (requiring an affirmative assent to terms) is enforceable\footnote{\href{https://blog.ericgoldman.org/archives/2022/12/hello-youve-been-referred-here-because-youre-wrong-about-web-scraping-laws-guest-blog-post-part-2-of-2.htm}{https://blog.ericgoldman.org/archives/2022/12/hello-youve-been-referred-here-because-youre-wrong-about-web-scraping-laws-guest-blog-post-part-2-of-2.htm}}.

The potential penalties are similar to breach of terms:
\begin{itemize}
    \item If you breach the ToS, the website operator may terminate your access to their services. 
    \item Violating the ToS might lead to civil lawsuits. For instance, if your actions cause financial harm to the website or other users, you could be held liable for damages.
    \item There have been cases where violating a website’s ToS was treated as a criminal offense. 
\end{itemize}

\subsection{Torts}

\subsubsection{Gross Negligence}

Gross negligence occurs in both common and statutory law. 

According to the Legal Information Institute, gross negligence is a lack of care that demonstrates reckless disregard for the safety or lives of others, which is so great it appears to be a conscious violation of other people's rights to safety. Gross negligence is a heightened degree of negligence representing an extreme departure from the ordinary standard of care. Falling between intent to do wrongful harm and ordinary negligence, gross negligence is defined as willful, wanton, and reckless conduct affecting the life or property or another. 

Potential Penalties
\begin{itemize}
    \item Civil penalties that vary by state
    \item A misdemeanor conviction for gross negligence can get you up to a year in jail. A felony conviction can result in several years of imprisonment.
    \item A felony conviction for gross negligence may also result in the loss of certain civil rights, such as the right to vote, serve on a jury, or possess a firearm.
\end{itemize}

\subsubsection{Defamation}

Defamation is a statement that injures a third party's reputation\footnote{\href{https://www.law.cornell.edu/wex/defamation}{https://www.law.cornell.edu/wex/defamation}}.

Elements:
\begin{enumerate}
    \item The defendant made a statement about the plaintiff to another party.
    \item The statement was intended to cause harm.
    \item The statement caused injury to the plaintiff's reputation.
    \item The statement was false.
    \item If the plaintiff is a public figure, then the defendant must have made the false statement intentionally or with reckless disregard. This is known as the actual malice standard. The same can be true if the plaintiff was not necessarily a public figure. They could instead be involved in some newsworthy event or a matter of public concern.
    \item The statement was not privileged information, such as any information shared between an attorney and a client or a doctor and a patient.
\end{enumerate}

Potential Penalties
\begin{itemize}
    \item Compensatory damages intended to compensate the victim of the defamation for the actual harm or loss they suffered as a result of the false statement. 
    \item Punitive damages intended to punish the person who made the false statement for their malicious or reckless conduct and deter them from engaging in similar behavior in the future.
    \item Injunctions 
    \item The court may order a retraction
\end{itemize}

\subsubsection{Public Disclosure of Private Facts}

Elements of the Tort:
\begin{enumerate}
    \item The Defendant gives publicity to a matter concerning the private life of another.
    \item The Plaintiff did not consent to the publication.
    \item The matter publicized or the act of publication would be highly offensive to a reasonable person.
    \item The publication was not of legitimate concern to the public.
\end{enumerate}

Penalties could include compensatory damages, punitive damages, and or injunctions. 

\subsubsection{False Light}

False light differs from defamation primarily in being intended "to protect the plaintiff's mental or emotional well-being," rather than to protect a plaintiff's reputation as is the case with the tort of defamation and in being about the impression created rather than being about veracity. If a publication of information is false, then a tort of defamation might have occurred. If that communication is not technically false but is still misleading, then a tort of false light might have occurred\footnote{\href{https://en.wikipedia.org/wiki/False\_light}{https://en.wikipedia.org/wiki/False\_light}}.

Elements of the Tort:
\begin{enumerate}
    \item The defendant published some information about the plaintiff.
    \item The information must portray the plaintiff in a false or misleading light.
    \item The information is highly offensive or embarrassing to a reasonable person.
    \item The defendant must have published the information with reckless disregard for its offensiveness.
\end{enumerate}

Penalties could include compensatory damages, punitive damages, and or injunctions. 

\subsubsection{Right to Publicity}

Personality rights are generally considered to consist of two types of rights: the right of publicity, or the right to keep one's image and likeness from being commercially exploited without permission or contractual compensation, which is similar (but not identical) to the use of a trademark; and the right to privacy, or the right to be left alone and not have one's personality represented publicly without permission. 

A commonly cited justification for this doctrine, from a policy standpoint, is the notion of natural rights and the idea that every individual should have a right to control how their right of publicity is commercialized by a third party, if at all\footnote{\href{https://en.wikipedia.org/wiki/Personality\_rights}{https://en.wikipedia.org/wiki/Personality\_rights}}.

Elements:
\begin{enumerate}
    \item Use of an individual's name or likeness 
    \item for commercial purposes
    \item without Plaintiff's consent
\end{enumerate}

Penalties could include compensatory damages, punitive damages, and or injunctions. 

\subsubsection{Appropriation of Name and Likeness}

Appropriation occurs when a defendant uses a plaintiff’s name, likeness, or image without their permission for commercial purposes. The right of publicity applies to individuals (for example, a celebrity) with a proven commercial value to their image or identity, while appropriation pertains to everyone.

Elements:
\begin{enumerate}
    \item The defendant used the plaintiff’s name or likeness, 
    \item the plaintiff did not consent, 
    \item the defendant gained a commercial benefit (or some other advantage), 
    \item the plaintiff was harmed, and 
    \item that the defendant’s conduct was a substantial factor in causing the plaintiff’s harm.
\end{enumerate}

Penalties could include compensatory damages, punitive damages, and or injunctions.  

\subsubsection{Failure to Warn}

The duty to warn arises in product liability cases, as manufacturers can be held liable for injuries caused by their products if the product causes an injury to a consumer and the manufacturer fails to supply adequate warnings about the risks of using the product or if they fail to supply adequate instructions for the proper use of the product. If the manufacturer fails to supply these warnings, the law will consider the product itself to be defective\footnote{\href{https://en.wikipedia.org/wiki/Duty\_to\_warn}{https://en.wikipedia.org/wiki/Duty\_to\_warn}}.

Elements:
\begin{enumerate}
    \item That the manufacturer knew of the danger posed by the product;
    \item That the manufacturer had a duty to warn consumers of the danger related to the product;
    \item The manufacturer was negligent in relation to their duty to warn; and
    \item The manufacturer’s failure to warn resulted in the plaintiff’s injury.
\end{enumerate}

Penalties could include compensatory damages, punitive damages, and or injunctions. 

\subsubsection{Libel}

Libel is a method of defamation expressed by print, writing, pictures, signs, effigies, or any communication embodied in physical form that is injurious to a person's reputation, exposes a person to public hatred, contempt or ridicule, or injures a person in his/her business or profession. State courts generally follow the common law of libel, which allows recovery of damages without proof of actual harm. Under the traditional rules of libel, injury is presumed from the fact of publication. However, the U.S. Supreme Court has held that the First Amendment's protection of freedom of expression limits a State's ability to award damages in actions for libel\footnote{\href{https://www.law.cornell.edu/wex/libel}{https://www.law.cornell.edu/wex/libel}}.

While it is sometimes said that the person making the libelous statement must have been intentional and malicious, actually it need only be obvious that the statement would do harm and is untrue. 

Elements:
\begin{enumerate}
    \item A defendant made a written factual and defamatory statement purporting to be fact;
    \item Regarding the plaintiff;
    \item That was published without privilege or authorization to others by the defendant; and
    \item There was resultant injury, or some harm caused to the reputation of the person or entity who is the subject of the statement, unless the statement falls within a category of “per se” harm.
\end{enumerate}

Penalties could include compensatory damages, punitive damages, and or injunctions. 

\subsection{Privacy}

\subsubsection{Children's Online Privacy Protection Act (COPPA)}

COPPA applies to the online collection of personal information by persons or entities under U.S. jurisdiction about children under 13 years of age, including children outside the U.S. if the website or service is U.S.-based. It details what a website operator must include in a privacy policy, when and how to seek verifiable consent from a parent or guardian, and what responsibilities an operator has to protect children's privacy and safety online, including restrictions on the marketing of those under 13.

Although children under 13 can legally give out personal information with their parents' permission, many websites—particularly social media sites, but also other sites that collect most personal info—disallow children under 13 from using their services altogether due to the cost and work involved in complying with the law.

Penalties could include fines up to \$51,744 per violation\footnote{\href{https://www.ftc.gov/business-guidance/resources/complying-coppa-frequently-asked-questions}{https://www.ftc.gov/business-guidance/resources/complying-coppa-frequently-asked-questions}}.

\subsubsection{GDPR: Right to Erasure}

The right to be forgotten (RTBF) is the right to have private information about a person be removed from Internet searches and other directories under some circumstances. The issue has arisen from desires of individuals to "determine the development of their life in an autonomous way, without being perpetually or periodically stigmatized as a consequence of a specific action performed in the past." It’s distinct from the right to privacy, which constitutes information that is not publicly known, whereas the right to be forgotten involves removing information that was publicly known at a certain time and not allowing third parties to access the information.

Penalties could be up to 4\% of global revenue. 

\section{Conclusion}

This taxonomy is the first paper to organize the many potential legal challenges generative AI creators may face. The stakes are high and the uncertainty is higher. While many laws may apply to the cutting edge of AI, it is unclear under what circumstances and to what extent the AI creator should be legally liable. 

Now that the claims have been identified, entities can better anticipate possible litigation claims and build stronger legal protections to ensure they are able to continue conducting the type of work they have decided to pursue. The taxonomy also helps to better identify some of the many open questions in law so researchers, policymakers, and legal scholars have a framework with which to prioritize and examine whether and how laws implemented decades ago can and should be applied to a type of artificial intelligence that sounded like science fiction just a few years ago. 

In addition to clarifying the bounds of the law, researchers, legal scholars, and practitioners could also consider possible solutions that are better suited for probabilistic tools like AI rather than the deterministic nature of tools in years past. Such solutions could include safe harbors so long as certain conditions are met to reduce harm or when conducting research. Another possible solution may be to broaden the application of the Learned Hand formula\footnote{No liability if the burden of adequate precautions is greater than likelihood of the harm times the magnitude of the harm if it occurs.} from torts to apply to most of the issues surrounding AI. This would be a nod to the fact that it’s impossible to anticipate all possible harms of AI, but it may be reasonable to require entities to implement best practices for their type of model given its capabilities.

For more information, Appendix ~\ref{lawsuits} contains a table of GenAI lawsuits, showing the parties involved and the claims made by the plaintiffs.

\bibliography{references}
\bibliographystyle{bib}

\newpage
\appendix

\section{Lawsuits Against GenAI Companies}
\label{lawsuits}

\begin{longtable}[c]{| p{0.125\linewidth} | p{0.125\linewidth} | p{0.53\linewidth} | p{0.1\linewidth} |}
\hline
\textbf{Plaintiff} & \textbf{Defendant} & \multicolumn{1}{c|}{\textbf{Claims}} & \textbf{Date filed} \\ \hline
J.L. & Alphabet, Deepmind, Google & \begin{enumerate}
    \item Violation Of California Unfair Competition Law, Business And Professions Code §§ 17200, Et Seq.
    \item Negligence
    \item Invasion Of Privacy Under California Constitution
    \item Intrusion Upon Seclusion5. Larceny/Receipt Of Stolen Property
    \item Conversion
    \item Unjust Enrichment
    \item Direct Copyright Infringement
    \item Vicarious Copyright Infringement
    \item Violation Of Digital Millennium Copyright Act, 17 U.S.C. § 1202(B)
\end{enumerate} & 7/11/2023 \\ \hline
Concord Music Group, Inc. et. al & Anthropic &
\begin{enumerate}
    \item Direct Copyright Infringement
    \item Contributory Infringement
    \item Vicarious Infringement
    \item Removal Or Alteration Of Copyright Management Information
\end{enumerate}
& 10/18/2023 \\ \hline
Michael Chabon & Meta &
\begin{enumerate}
    \item Direct Copyright Infringement, 17 U.S.C. § 106, Et Seq.
    \item Vicarious Copyright Infringement 17 U.S.C. § 106
    \item Digital Millennium Copyright Act – Removal Of Copyright Management Information 17 U.S.C. § 1202(B)
    \item Violations Of The California Unfair Competition Law Cal. Bus. \& Prof. Code §§ 17200, Et Seq
    \item Negligence
    \item Unjust Enrichment
\end{enumerate} & 9/12/2023 \\ \hline
Mike Huckabee & Meta, Bloomberg, Microsoft, The EleutherAI Institute &
\begin{enumerate}
    \item Direct Copyright Infringement, 17 U.S.C. § 106, Et Seq.
    \item Vicarious Copyright Infringement, 17 U.S.C. § 106
    \item Digital Millennium Copyright Act – Removal Of Copyright Management Information, 17 U.S.C. § 1202(B)
    \item Conversion
    \item Negligence
    \item Unjust Enrichment
\end{enumerate} & 10/17/2023 \\ \hline
Richard Kadrey & Meta &
\begin{enumerate}
    \item Direct Copyright Infringement 17 U.S.C. § 106
    \item Vicarious Copyright Infringement 17 U.S.C. § 106
    \item Removal Of Copyright-Management Information And False Assertion Of Copyright 17 U.S.C. § 1202(B) 
    \item Unfair Competition Cal. Bus. \& Prof. Code §§ 17200, Et Seq. 
    \item Unjust Enrichment California Common Law
    \item Negligence California Common Law
\end{enumerate}& 7/7/2023\\ \hline
Abdi Nazemian & NVIDIA & \begin{enumerate}
    \item Direct Copyright Infringement (17 U.S.C. § 501)
\end{enumerate} & 3/8/2024\\ \hline
Authors Guild & OpenAI &
\begin{enumerate}
    \item Direct Copyright Infringement (17 U.S.C. § 501) 
    \item Vicarious Copyright Infringement
    \item Contributory Copyright Infringement
\end{enumerate}& 9/19/2023  \\ \hline
The New York Times & Microsoft, OpenAI &
\begin{enumerate}
    \item Copyright Infringement (17 U.S.C. § 501)
    \item Vicarious Copyright Infringement
    \item Contributory Copyright Infringement
    \item Digital Millennium Copyright Act – Removal Of Copyright Management Information (17 U.S.C. § 1202)
    \item Common Law Unfair Competition By Misappropriation
    \item Trademark Dilution (15 U.S.C. § 1125(C)) 
\end{enumerate} & 12/27/2023 \\ \hline
Julian Sancton/Alter & OpenAI, Microsoft &
\begin{enumerate}
    \item Copyright Infringement (17 U.S.C. § 501)
    \item Contributory Infringement
\end{enumerate} & 11/21/2023 \\ \hline
A.T., J.H.  & OpenAI, Microsoft &
\begin{enumerate}
    \item Violation Of Electronic Communications Privacy Act, 18 U.S.C. §§ 2510, Et Seq. 
    \item Violation Of The Computer Fraud And Abuse Act, 18 U.S.C. § 1030
    \item Violation Of The California Invasion Of Privacy Act (“Cipa”), Cal. Penal Code § 631
    \item Violation Of California Unfair Competition Law, Business And Professions Code §§ 17200, Et Seq.
    \item Negligence
    \item Invasion Of Privacy
    \item Intrusion Upon Seclusion
    \item Larceny/Receipt Of Stolen Property
    \item Conversion
    \item Unjust Enrichment
    \item New York General Business Law §§ 349, Et Seq. 
\end{enumerate} & 9/5/2023\\ \hline
Basbanes  & Microsoft, OpenAI &
\begin{enumerate}
    \item Direct Copyright Infringement (17 U.S.C. § 501)
    \item Vicarious Copyright Infringement
    \item Contributory Copyright Infringement
\end{enumerate} & 1/5/2023\\ \hline
The Intercept Media  & OpenAI, Microsoft &
\begin{enumerate}
    \item Violation of 17 U.S.C. § 1202(b)(1)
    \item Violation of 17 U.S.C. § 1202(b)(3)
\end{enumerate} & 2/28/2024  \\ \hline
Raw Story Media  & OpenAI & \begin{enumerate}
    \item Violation of 17 U.S.C. § 1202(b)(1)
\end{enumerate}  & 2/28/2024  \\ \hline
A.S. & OpenAI, Microsoft &
\begin{enumerate}
    \item Violation Of Electronic Communications Privacy Act, 18 U.S.C. §§ 2510, Et Seq. 
    \item Violation Of The Computer Fraud And Abuse Act, 18 U.S.C. § 1030
    \item Violation Of The California Invasion Of Privacy Act (“Cipa”), Cal. Penal Code § 631
    \item Violation Of California Unfair Competition Law, Business And Professions Code §§ 17200, Et Seq.
    \item Negligence
    \item Invasion Of Privacy
    \item Intrusion Upon Seclusion
    \item Larceny/Receipt Of Stolen Property
    \item Conversion
    \item Unjust Enrichment
\end{enumerate} & 2/27/2024  \\ \hline
Michael Chabon & OpenAI &
\begin{enumerate}
    \item Direct Copyright Infringement, 17 U.S.C. § 106, Et Seq. 
    \item Vicarious Copyright Infringement 17 U.S.C. § 106
    \item Digital Millennium Copyright Act – Removal Of Copyright Management Information 17 U.S.C. § 1202(B)
    \item Violations Of The California Unfair Competition Law Cal. Bus. \& Prof. Code §§ 17200, Et Seq.
    \item Negligence
    \item Unjust Enrichment
\end{enumerate} & 9/8/2023\\ \hline
Sarah Silverman  & OpenAI &
\begin{enumerate}
    \item Direct Copyright Infringement 17 U.S.C. § 106
    \item Vicarious Copyright Infringement 17 U.S.C. § 106
    \item Digital Millennium Copyright Act— Removal Of Copyright Management Information 17 U.S.C. § 1202(B)
    \item Unfair Competition Cal. Bus. \& Prof. Code §§ 17200, Et Seq
    \item Negligence Under California Common Law
    \item Unjust Enrichment Under California Common Law
\end{enumerate} & 7/7/2023\\ \hline
Paul Tremblay & OpenAI &
\begin{enumerate}
    \item Direct Copyright Infringement 17 U.S.C. § 106
    \item Vicarious Copyright Infringement 17 U.S.C. § 106
    \item Digital Millennium Copyright Act—Removal Of Copyright Management Information 17 U.S.C. § 1202(B)
    \item Unfair Competition Cal. Bus. \& Prof. Code §§ 17200, Et Seq.
    \item Negligence
    \item Unjust Enrichment
\end{enumerate} & 6/28/2023  \\ \hline
Mark Walters & OpenAI &
\begin{enumerate}
    \item Negligence
    \item Libel
\end{enumerate}
& 6/5/2023\\ \hline
Doe 1  & OpenAI, Github, Microsoft  &
\begin{enumerate}
    \item Violation Of The Digital Millennium Copyright Act 17 U.S.C. §§ 1201–1205
    \item Breach Of Contract—Open-Source License Violations Common Law
    \item Tortious Interference In A Contractual Relationship Common Law
    \item Fraud Common Law
    \item False Designation Of Origin—Reverse Passing Off 15 U.S.C. § 1125
    \item Unjust Enrichment Cal. Bus. \& Prof. Code §§ 17200, Et Seq. And Common Law
    \item Unfair Competition 15 U.S.C. § 1125; Cal. Bus. \& Prof. Code §§ 17200, Et Seq.; And Common Law
    \item Breach Of Contract Violation Of Github Privacy Policy And Terms Of Service Cal. Bus. \& Prof. Code § 22575–22579; Cal. Civ. Code § 1798.150; And Common Law
    \item Violation Of The California Consumer Privacy Act Cal. Civ. Code § 1798.150
    \item Negligence—Negligent Handling Of Personal Data Common Law
    \item Civil Conspiracy Common Law
    \item Declaratory Relief 28 U.S.C. § 2201(A) And Cal. Code Civ. Proc. § 1060
\end{enumerate} & 11/3/2022  \\ \hline
Thomson Reuters  & ROSS Intelligence  &
\begin{enumerate}
    \item Copyright Infringement (17 U.S.C. § 101 et seq.)
    \item Tortious Interference with Contract
\end{enumerate}
& 9/25/2023  \\ \hline
Sarah Andersen & Stability AI, Midjourney, DeviantArt  &
\begin{enumerate}
    \item Direct Copyright Infringement 17 U.S.C. §§ 106, Et Seq.
    \item Vicarious Copyright Infringement 17 U.S.C. §§ 106, Et Seq.
    \item Violation Of The Digital Millennium Copyright Act 17 U.S.C. §§ 1201–1205
    \item Violation Of The Statutory Right Of Publicity Cal. Civ. Code § 3344
    \item Violation Of The Common Law Right Of Publicity Common Law
    \item Unfair Competition 15 U.S.C. § 1125; Cal. Bus. \& Prof. Code §§ 17200, Et Seq.; And Common Law
    \item Breach Of Contract Violation Of Deviantart Policies Cal. Bus. \& Prof. Code § 22575–22579; Cal. Civ. Code § 1798.150; And Common Law
    \item Declaratory Relief 28 U.S.C. § 2201(A) And Cal. Code Civ. Proc. § 1060
\end{enumerate} & 1/13/2023  \\ \hline
Getty Images  & Stability AI &
\begin{enumerate}
    \item Copyright Infringement (17 U.S.C. § 101 Et Seq.)
    \item Providing False Copyright Management Information In Violation Of 17 U.S.C. § 1202(A)
    \item Removal Or Alteration Of Copyright Management Information In Violation Of Section 1202(B)
    \item Trademark Infringement In Violation Of Section 32 Of The Lanham Act, 15 U.S.C. § 1114(1)
    \item Unfair Competition In Violation Of Section 43(A) Of The Lanham Act, 15 U.S.C. § 1125(A)
    \item Trademark Dilution In Violation Of Section 43(C) Of The Lanham Act, 15 U.S.C. § 1125(C)
    \item Deceptive Trade Practices In Violation Of Delaware’s Uniform Deceptive Trade Practices Act
    \item Trademark Dilution In Violation Of Section 3313 Of The Delaware Trademark Act
\end{enumerate} & 2/3/2023 \\ \hline
 Paul Lehrman and Linnea Sage & Lovo &
\begin{enumerate}
    \item Violation of New York Civil Rights Law Sections 50, 51
    \item Deceptive Acts and Practices in Violation of the New York Deceptive Practices Act, N.Y. GBL § 349
    \item False advertising in violation of the New York False Advertising Act, N.Y. GBL § 350
    \item Unfair Competition and False Affiliation in Violation of Section 43 of the Lanham Act, 15 U.S.C. § 1125(a)
    \item False Advertising in Violation of Section 43 of the Lanham Act, 15 U.S.C. § 1125(a)
    \item Unjust Enrichment
    \item Tortious Interference with Advantageous Business Relationship
    \item Fraud
\end{enumerate} & 5/16/2024 \\ \hline
 Dubus et al & Nvidia &
\begin{enumerate}
    \item Direct Copyright Infringement (17 U.S.C. § 501)
\end{enumerate} & 5/2/2024 \\ \hline
 Rebecca Makkai and Jason Reynolds & Databricks and Mosaic &
\begin{enumerate}
    \item Direct Copyright Infringement (17 U.S.C. § 501)
    \item Vicarious Copyright Infringement
\end{enumerate} & 5/2/2024 \\ \hline
 Jingna Zhang, Sarah Andersen,
Hope Larson, and
Jessica Fink & Google &
\begin{enumerate}
    \item Direct Copyright Infringement (17 U.S.C. § 501)
    \item Vicarious Copyright Infringement
\end{enumerate} & 4/30/2024 \\ \hline
Daily News, Chicago Tribune Company,
Orlando Sentinel, 
Sun-Sentinel, San Jose MercuryNews, DP Media Network, ORB Publishing, and Northwest Publications & Microsoft, OpenAI &
\begin{enumerate}
    \item Copyright Infringement (17 U.S.C. § 501)
    \item Vicarious Copyright Infringement
    \item Contributory Copyright Infringement
    \item Digital Millennium Copyright Act – Removal of Copyright Management Information (17 U.S.C. § 1202)
    \item Common Law Unfair Competition By Misappropriation
    \item Trademark Dilution (15 U.S.C. § 1125(c))
    \item Dilution and Injury to Business Reputation (N.Y. Gen. Bus. Law § 360-l)
\end{enumerate} & 4/30/2024 \\ \hline
 Nazemian et al & Nvidia &
\begin{enumerate}
    \item Direct Copyright Infringement (17 U.S.C. § 501)
\end{enumerate} & 3/8/2024 \\ \hline
 Stewart O’Nan,
Abdi Nazemian, and
Brian Keene & Databricks and MosaicML &
\begin{enumerate}
    \item Direct Copyright Infringement (17 U.S.C. § 501)
    \item Vicarious Copyright Infringement
\end{enumerate} & 3/8/2024 \\ \hline
\caption{Lawsuits against GenAI companies.}
\label{tab:ongoing-lawsuits}
\end{longtable}


\end{document}